\documentclass[11pt,twoside]{article}  
\usepackage{apn3conf}

\begin{document}   

%
%
%
%

\title{Hot Gas in Planetary Nebulae}

%
%
%

\author{You-Hua Chu, Robert A.\ Gruendl, Mart\'\i n A.\ Guerrero}
\affil{Astronomy Department, University of Illinois, 1002 W. Green St.,
  Urbana, IL 61801}

%
%

\contact{You-Hua Chu}
\email{chu@astro.uiuc.edu}

%
%
%
%
%

\paindex{Chu, Y.-H.}
\aindex{Gruendl, R.\ A.}
\aindex{Guerrero, M.\ A.}

%
%

\authormark{Chu,Gruendl, \& Guerrero}

%
%



\begin{abstract}          
Diffuse X-ray emission has been detected in a small number of 
planetary nebulae (PNe), indicating the existence of shocked
fast stellar winds and providing support for the 
interacting-stellar-winds formation scenario of PNe.
However, the observed X-ray luminosities are much lower than
expected, similar to the situation seen in bubbles or superbubbles
blown by massive stars.
Ad hoc assumptions have been made to reconcile the discrepancy
between observations and theoretical expectations.
We have initiated {\sl FUSE} programs to observe O\,VI absorption
and emission from PNe, and our preliminary results indicate that 
O\,VI emission provides an effective diagnostic for hot gas in PN 
interiors.

\end{abstract}

%
%

\section{Introduction}
The formation of planetary nebulae (PNe) is a complex issue.
The current fast stellar wind from the central star of a PN must 
interact with the previous slow AGB wind; it has thus been suggested 
that PNe are formed by interacting stellar winds (Kwok, Purton, \&
Fitzgerald 1978).
The frequently observed multi-polar and point-symmetric structures 
of young PNe suggest that collimated outflows are responsible for 
shaping PNe in early evolutionary stages (Sahai \& Trauger 1998).
The fast stellar winds and high-velocity collimated outflows, upon
impacting the ambient medium, will be shocked to a temperature of
$T = (3/16)\mu v^2 /k$, where $\mu$ is the average mass per particle,
$v$ is the shock velocity, and $k$ is the Boltzmann constant.
Hot gas at X-ray-emitting temperatures, i.e., $\ge1\times10^6$ K, 
can be generated for shock velocities greater than $\sim$300 
km~s$^{-1}$.

As reviewed by Guerrero et al.\ (in this volume), X-ray observations
have revealed hot gas in a number of young PNe such as NGC\,6543 and
NGC\,7027, and established stringent upper limits on the hot gas 
content in evolved PNe such as NGC\,7293 (the Helix).
The detection of diffuse X-ray emission from PNe provides a strong 
support for the interacting-stellar-winds scenario of PN formation.
This formation mechanism is in fact similar to that for circumstellar
bubbles blown by Wolf-Rayet (WR) stars.
Massive WR stars are descendants of red supergiants or luminous blue
variables which lose mass via copious slow winds; the fast WR wind
sweeps up the previous slow wind and forms a circumstellar bubble.
Comparisons between PNe and WR bubbles will help us understand both
types of objects.
In the first part of this paper, we will discuss the physical 
properties of the 10$^6$~K hot gas in PN interiors and make 
comparisons with shocked stellar winds in WR bubbles.

In the second part of this paper, we will discuss hot gas at 
cooler temperatures, i.e., a few $\times10^5$~K, in order to:
(1) determine whether some evolved PNe are filled with such
    cooler hot gas, and
(2) study the interface layer between the 10$^6$~K interior gas 
    and the 10$^4$~K nebular shell in young PNe.
We will present far-UV observations of the 10$^5$~K gas in PNe
and use the results to make predictions about the presence or
absence of 10$^6$~K hot gas in PN interiors.

\section{$10^6$~K Hot Gas}
{\sl Chandra} and {\sl XMM-Newton} observations of diffuse X-ray
emission from PNe allow us to examine the spatial distribution
and temperatures of the hot gas.
Among the six PNe with diffuse X-ray emission, Mz\,3 and NGC\,6543 
are the only two that are adequately resolved by the instrumental
point spread function, and both of them show a limb-brightened 
X-ray morphology, indicating that the dense X-ray-emitting gas is 
concentrated near the inner walls of their nebular shells.
This is similar to what we have observed in the WR bubbles NGC\,6888 
and S\,308 (Chu et al.\ 2003; Gruendl et al.\ 2004); see Fig.~1 for 
an {\sl XMM-Newton} EPIC image of S\,308 and a comparison with 
an optical image.
The large angular size of S\,308 makes it possible to see the
detailed relationship between the hot interior gas and the cooler
nebular shell.

\begin{figure}[b]
\plottwo{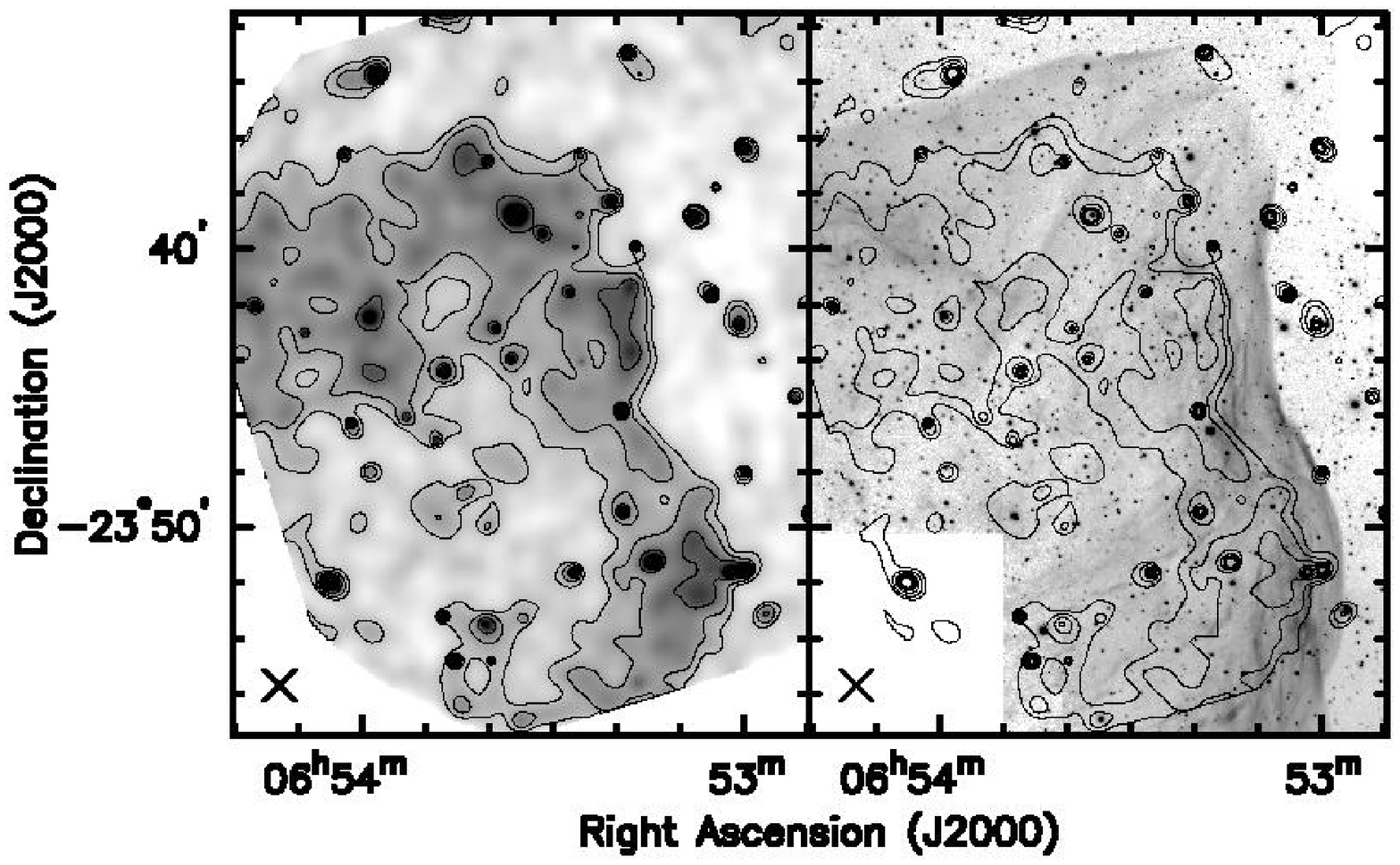}{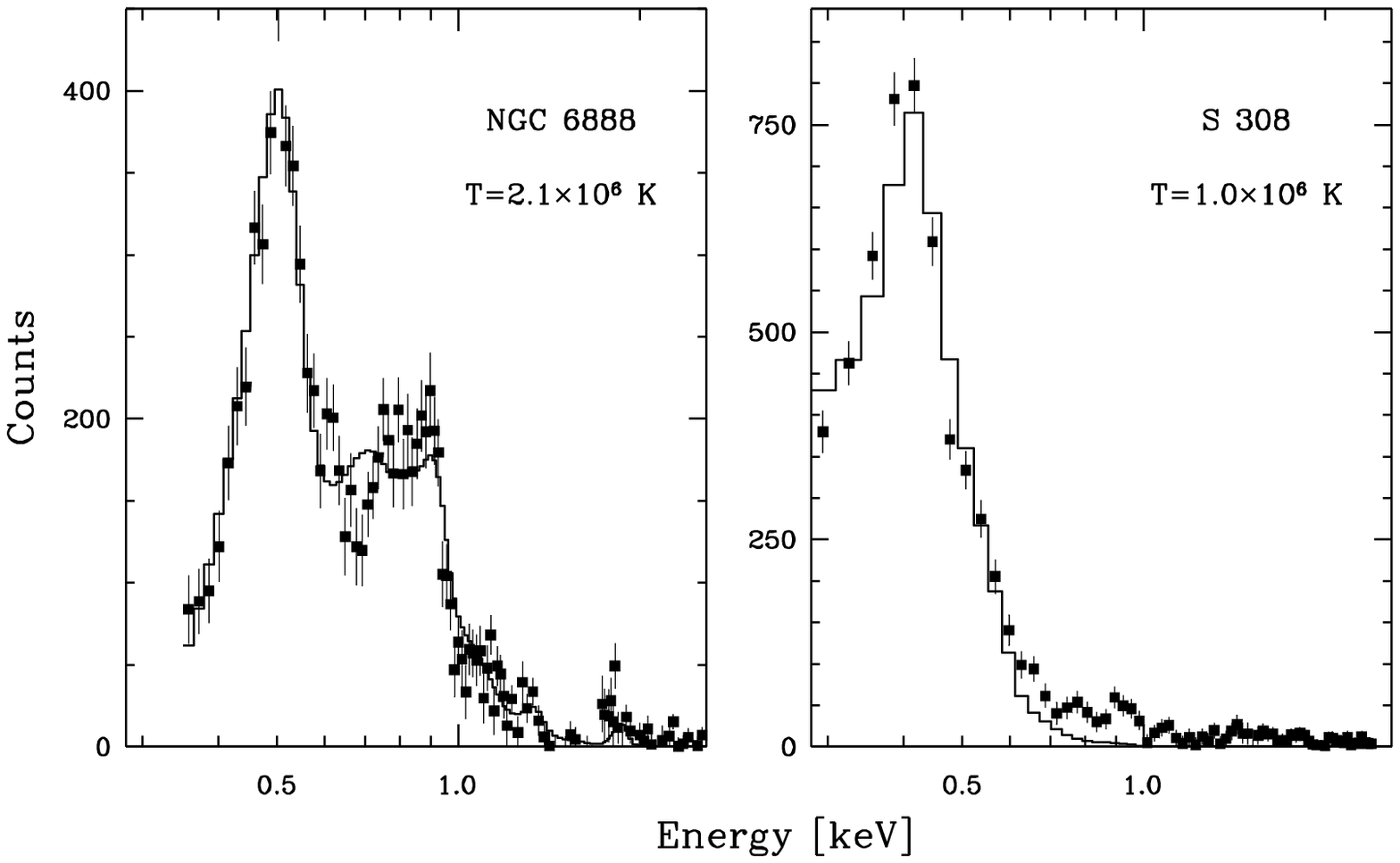}
\caption{{\sl Left}: {\sl XMM-Newton} EPIC image of the NW 
quadrant of the WR bubble S\,308, and an [O\,III]$\lambda$5007 line
image overlaid with X-ray contours.  The location of the central WR
star is marked with an ``$\times$''.  
{\sl Right}: {\sl Chandra} ACIS-S spectrum of the hot gas in
NGC\,6888 and {\sl XMM-Newton} EPIC/pn spectrum of the hot gas 
in S\,308.}
\end{figure}

X-ray spectra of the hot gas in PN interiors (Guerrero et al.\ in 
this volume) show a narrow range of temperatures, $2-3\times10^6$~K.
These temperatures are much lower than the $\sim10^7$~K gas detected
in superbubbles around young clusters, where colliding stellar winds 
may have produced the hot gas and the high temperatures (Townsley et 
al.\ 2003).
On the other hand, the temperatures of the X-ray-emitting gas in PNe
are similar to the $1-2\times10^6$~K gas detected in the WR bubbles 
NGC\,6888 and S\,308 (see their spectra in Fig.~1).

Both PNe and WR bubbles have been modeled hydrodynamically
(Zhekov \& Perinotto 1996, 1998; Garc\'\i a-Segura et al.\
1996a,b), assuming that the fast stellar wind is shock-heated 
to high temperatures and forms a contact discontinuity with 
the photoionized 10$^4$~K nebular shell.
It is further assumed that thermal conduction and mass evaporation
take place at the hot/cold gas interface, which lowers the 
temperature and raises the density of the hot interior gas.
In fact, the hot interior gas is dominated by the evaporated
nebular material.

As the X-ray emission is proportional to the square of the density, 
the integrated X-ray spectrum of the hot gas will be dominated by the
emission from the densest gas which is at the lowest temperatures.
Therefore, the low temperatures derived from model fits to the X-ray
spectra are expected.
The X-ray luminosities of the diffuse emission from PNe and WR bubbles
are, however, at least an order of magnitude lower than those
expected from models (Chu et al.\ 2001; Wrigge et al.\ 1994), implying 
that the amount of hot gas is significantly less than that predicted 
by models with thermal conduction.

Several possibilities have been suggested to reconcile the 
discrepancies between observation and theory (Arnaud et al.\
1996; Soker \& Kastner 2003):
(1) the fast stellar wind may have evolved with time, with a lower 
    velocity or mass loss rate in the past;
(2) the fast stellar wind is anisotropic, with a bipolar structure; 
    and
(3) the thermal conduction at the hot/cold gas interface may be
    inhibited by the presence of magnetic fields.
Unfortunately, diffuse X-ray emission has been detected in only six
PN interiors; two PNe are not adequately resolved spatially, and
another two are detected with less than 100 counts.
The current X-ray observations are clearly inadequate to constrain
models and to discern the above possibilities.
High-quality X-ray observations of PNe are badly~needed.

\section{10$^5$~K Hot Gas}

Hot gas at 10$^5$~K is traditionally studied with spectral lines
of highly ionized species, such as C\,IV, N\,V, and O\,VI.
If these ions are produced by thermal collisions, C\,IV, N\,V,
and O\,VI will be the dominant ionization stages for C, N, and
O at temperatures of $1\times10^5$, $2\times10^5$, and 
$3\times10^5$ K, respectively.
As the ionization potentials of C\,III, N\,IV, and O\,V are
47.9, 77.5, and 113.9 eV, photoionization of these ions
becomes significant in PNe with central stars at stellar 
effective temperatures greater than $\sim$ 35,000, 75,000,
and 125,000 K, respectively.
Therefore, O\,VI provides the best diagnostic for 10$^5$ K gas,
as it is the least easily confused by photoionization.

The O\,VI $\lambda\lambda$1032,1037 lines can be observed with 
the recently launched {\sl Far Ultraviolet Spectroscopic Explorer
(FUSE)}.
Four {\sl FUSE} observing programs were 
carried out to search for 10$^5$~K gas in PN interiors.
Some observations use spectra of the central stars to search for
nebular O\,VI absorption, while others use spectra of the nebula
(avoiding the central star) to search for O\,VI emission.
These four programs are:
\begin{itemize}
\item O\,VI absorption observation of 10 PNe (PI: Gruendl; AO2)
\item O\,VI emission observation of NGC\,6543 (PI: Guerrero; AO3)
\item O\,VI emission observation of NGC\,7009 (PI: Iping; AO3)
\item O\,VI emission observation of 9 PNe (PI: Guerrero; AO4)
\end{itemize}

\begin{figure}[t]
\epsscale{0.6}
\plotone{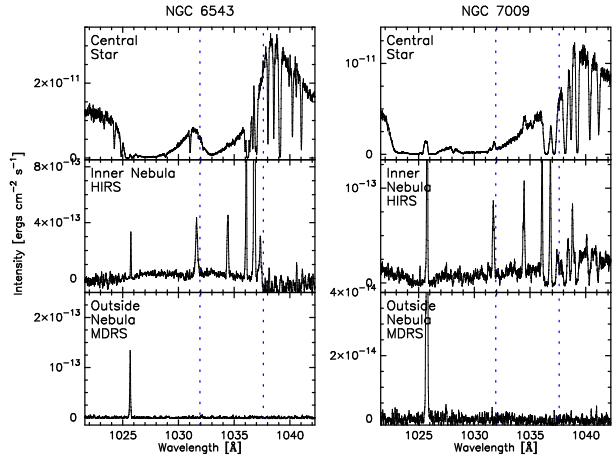}
\caption{{\sl FUSE} spectra of NGC\,6543 and NGC\,7009, two 
PNe with known diffuse X-ray emission. The vertical dashed lines
mark the positions of the O\,VI lines. See text for more 
descriptions.}
\end{figure}

These observations have produced many exciting results.
First of all, {\sl FUSE} observations detected O\,VI in
NGC\,6543 and NGC\,7009, two PNe with known diffuse X-ray 
emission.  
The top panels of Fig.~2 display {\sl FUSE} spectra of
their central stars; both show pronounced P Cygni profiles
in the stellar O\,VI lines, indicating the presence of 
fast stellar wind.
NGC\,6543 shows a stellar spectrum with nebular O\,VI absorption 
at a velocity blue-shifted from the nebular systemic velocity,
indicating that it is associated with the interface layer
on the approaching side of the nebular shell.  
NGC\,7009 does not show obvious nebular O\,VI absorption in 
its stellar spectrum.
The middle panels of Fig.~2 display {\sl FUSE} spectra of 
apertures centered inside the nebulae.
Both nebulae show obvious O\,VI emission lines.
The bottom panels of Fig.~2 display {\sl FUSE} spectra of
apertures centered outside the nebulae, where only airglow
lines are seen.
{\sl FUSE} observations of NGC\,6543 and NGC\,7009 demonstrate
clearly that the O\,VI emission from interfaces between 
the 10$^6$~K interior and the 10$^4$~K nebular shell can
be detected.
Quantitative analysis of the O\,VI emission and absorption 
lines, complicated by the interstellar and circumstellar 
H$_2$ absorption along the line of sight, is in progress.

\begin{figure}[t]
\epsscale{0.6}
\plotone{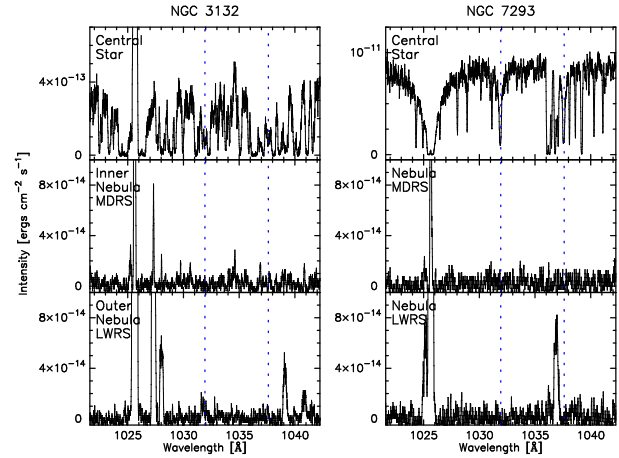}
\caption{{\sl FUSE} spectra of NGC\,3132 and NGC\,7293.
Note the contrasting stellar and nebular O\,VI lines between 
these spectra and those in Fig.~2. The vertical dashed lines
mark the positions of the O\,VI lines.  See text for more
descriptions.}
\end{figure}

{\sl FUSE} observations of the Helix Nebula (Fig.~3), a PN 
confirmed to lack 10$^6$~K gas, show contrasting spectra in both the
central star and the nebula.
The spectrum of the central star shows stellar O\,VI absorption,
but not P Cygni profiles, indicating a lack of fast stellar 
wind, which is consistent with the conclusion derived from 
{\sl IUE} observations (Cerruti-Sola \& Perinotto 1985).
Neither O\,VI absorption nor O\,VI emission from the nebula
is detected.
It is thus reasonable to expect that a lack of both stellar 
P Cygni profile in O\,VI and nebular O\,VI is a good indication 
of an absence of hot gas in the PN interior.
NGC\,3132 is an awarded {\sl Chandra} target in Cycle 5.
The {\sl FUSE} observations of its central star and the nebula,
displayed in Fig.~3, show neither stellar P Cygni O\,VI nor
nebular O\,VI.
We predict that no 10$^6$~K hot gas exists in NGC\,3132 and 
that {\sl Chandra} observations of NGC\,3132 will not detect 
any diffuse X-ray emission.

\section{Future Work}
X-ray observations have shown that hot gas exists in some 
PN interiors, and this interior hot gas may play an important 
role in the formation and evolution of PNe.
To understand the interacting stellar winds in PNe, high S/N
and high spatial resolution X-ray observations of more PNe 
are needed.
{\sl FUSE} observations of O\,VI emission offer an effective
diagnostic for the presence of hot gas in PNe, and should be
used to guide the target selection for X-ray observations.
Future models of interacting stellar winds need to consider
the plasma micro-physics and dynamical mixing of nebular
material, in addition to a critical assessment of the
degree of thermal conduction.
When high-quality X-ray spectra are available, they need to
be modeled with considerations of non-equilibrium ionization 
and temperature structure in the hot gas.

%
%
%
%


\end{document}